\documentclass[sigconf]{acmart}
\AtBeginDocument{%
  }

\setcopyright{acmlicensed}
\copyrightyear{2026}
\acmYear{2026}
\acmDOI{XXXXXXX.XXXXXXX}
\acmConference[Conference acronym 'XX]{Make sure to enter the correct
  conference title from your rights confirmation email}{June 03--05,
  2018}{Woodstock, NY}
\acmISBN{978-1-4503-XXXX-X/2018/06}

\usepackage{algorithmic}
\usepackage{algorithm}
\usepackage{bibentry}
\usepackage{multirow}
\usepackage{xcolor}
\usepackage{colortbl}
\usepackage{enumitem}
\usepackage{CJKutf8}
\usepackage{subfig}

\usepackage{graphicx}
\usepackage{stfloats}    
\usepackage{afterpage}    
\usepackage{lipsum}      
\usepackage{kantlipsum}   
\usepackage{makecell}

\newcommand{\modelname}{Promise}

\begin{document}
\title{PROMISE: Process Reward Models Unlock Test-Time Scaling Laws in Generative Recommendations}


\author{Chengcheng Guo$^\text{*}$}
\email{guochengcheng03@kuaishou.com}
\affiliation{%
  \institution{Kuaishou Inc.}
  \city{Beijing}
  \country{China}
}

\author{Kuo Cai$^\text{*}$}
\email{caikuo@kuaishou.com}
\affiliation{%
  \institution{Kuaishou Inc.}
  \city{Beijing}
  \country{China}
}

\author{Yu Zhou}
\email{zhouyu12@kuaishou.com}
\affiliation{%
  \institution{Kuaishou Inc.}
  \city{Beijing}
  \country{China}
}

\author{Qiang Luo$^{\dag}$}
\email{luoqiang@kuaishou.com}
\affiliation{%
  \institution{Kuaishou Inc.}
  \city{Beijing}
  \country{China}
}

\author{Ruiming Tang$^{\dag}$}
\email{tangruiming@kuaishou.com}
\affiliation{%
  \institution{Kuaishou Inc.}
  \city{Beijing}
  \country{China}
}

\author{Han Li}
\email{lihan08@kuaishou.com}
\affiliation{%
  \institution{Kuaishou Inc.}
  \city{Beijing}
  \country{China}
}

\author{Kun Gai}
\email{gai.kun@qq.com}
\affiliation{%
  \institution{Unaffiliated}
  \city{Beijing}
  \country{China}
}

\author{Guorui Zhou}
\email{zhouguorui@kuaishou.com}
\affiliation{%
  \institution{Kuaishou Inc.}
  \city{Beijing}
  \country{China}
}

\thanks{$^\text{*}$ Equal contribution}
\thanks{$^{\dag}$ Corresponding author}
\renewcommand{\shortauthors}{Guo et al.}

\begin{abstract}

Generative Recommendation has emerged as a promising paradigm, reformulating recommendation as a sequence-to-sequence generation task over hierarchical Semantic IDs. However, existing methods suffer from a critical issue we term Semantic Drift, where errors in early, high-level tokens irreversibly divert the generation trajectory into irrelevant semantic subspaces. Inspired by Process Reward Models (PRMs) that enhance reasoning in Large Language Models, we propose \textbf{\textsc{\modelname}}, a novel framework that integrates dense, step-by-step verification into generative models. \textsc{\modelname} features a lightweight PRM to assess the quality of intermediate inference steps, coupled with a PRM-guided Beam Search strategy that leverages dense feedback to dynamically prune erroneous branches. Crucially, our approach unlocks Test-Time Scaling Laws for recommender systems: by increasing inference compute, smaller models can match or surpass larger models. Extensive offline experiments and online A/B tests on a large-scale platform demonstrate that \textsc{\modelname} effectively mitigates Semantic Drift, significantly improving recommendation accuracy while enabling efficient deployment.
\end{abstract}



\received{8 January 2026}
\received[revised]{8 January 2026}
\received[accepted]{8 January 2026}

\maketitle

\section{Introduction}

Recommender Systems (RS) have witnessed a fundamental shift from the conventional "Retrieve-then-Rank" pipeline to Generative Recommendation \cite{202512.0203}. Inspired by Large Language Models (LLMs) \cite{achiam2023gpt,liu2024deepseek}, this paradigm reformulates recommendation as a sequence-to-sequence generation task \cite{rajput2023recommender}. By directly generating the next item identifier, these models facilitate end-to-end learning of user preferences. They demonstrate superior modeling precision and remarkable scaling capabilities \cite{zhou2025onerectechnicalreport,zhou2025onerecv2}. Consequently, this approach has rapidly emerged as a mainstream direction, with applications spanning from e-commerce to short-video platforms \cite{deng2025onerec,wei2025oneloc,chen2025onesearch}.

A key driver of this success is the adoption of Semantic IDs (SIDs) \cite{singh2024better}. Unlike random atomic IDs, Semantic IDs are discrete token sequences derived from hierarchical quantization methods, such as RQ-VAE \cite{lee2022autoregressive} or Residual K-means \cite{luo2025qarm}. Leading approaches like TIGER \cite{rajput2023recommender}, LC-Rec \cite{zheng2024adapting}, and One-series models \cite{deng2025onerec,wei2025oneloc} leverage these tokens to represent items for generative retrieval. However, these methods suffer from a critical issue that we formally define as \textbf{Semantic Drift}, where the generation trajectory gradually deviates from the user's true intent due to error accumulation \cite{bengio2015scheduled,shi2025llada}. This phenomenon is rooted in Exposure Bias—the discrepancy between training and inference. During training, models typically employ Teacher Forcing, where the next token is predicted conditioned on the ground-truth history. In contrast, during inference, the model must operate autoregressively, conditioning on its own previously generated tokens. This mismatch means the model is never exposed to its own prediction errors during training. Consequently, when a deviation occurs at inference time—especially in high-level semantic tokens—the model lacks the capability to recover, causing the generation to drift into an irrelevant semantic subspace (e.g., from "Electronics" to "Home Appliances").

\begin{figure}
    \centering
    \includegraphics[width=\linewidth]{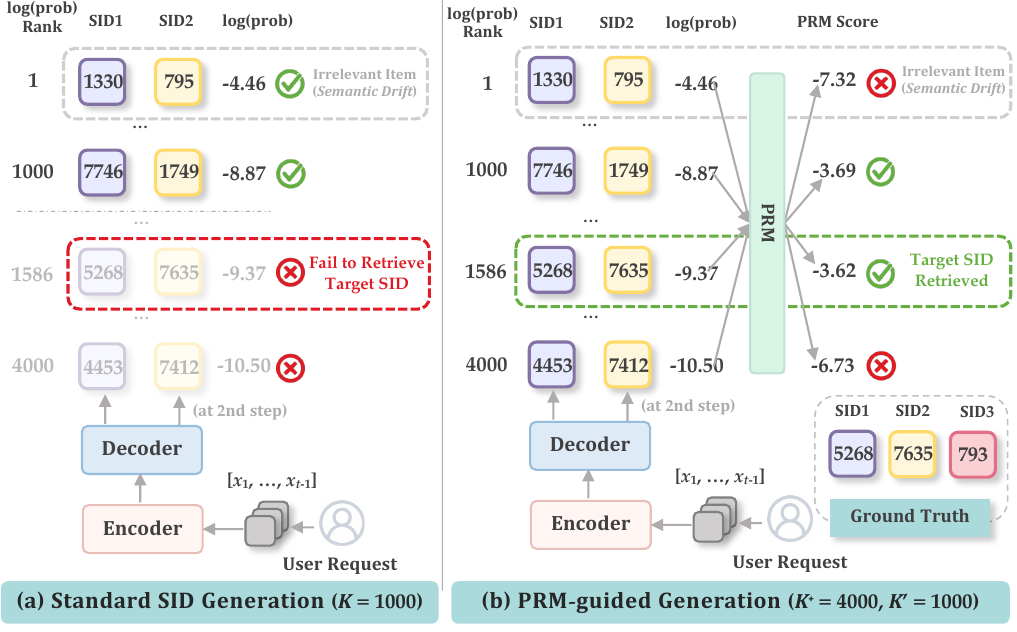}
    \caption{A specific case of SID generation based on the same user $u$ and context $c$ using standard beam search and PRM-guided beam search, respectively.  
    (a): We observe the semantic drift where the model generates irrelevant SIDs, failing to retrieve target SIDs. (b): With PRM-guided search, the PRM evaluates and ranks the generation quality. The model successfully selects target SIDs and eliminates erroneous outputs caused by semantic drift.}\vspace{-0.6cm}
    \label{fig:intro}
\end{figure}

A parallel challenge is observed in LLMs, particularly within complex mathematical reasoning and code generation tasks \cite{wei2022chain,chen2021evaluating}. In these domains, a single logical error in an intermediate step often propagates, rendering the final solution incorrect \cite{dziri2023faith}. Recent breakthroughs demonstrate that \textbf{Process Reward Models (PRMs)} \cite{lightman2023let}—which evaluate the correctness of intermediate reasoning steps—can effectively address the sparse signal limitations of Outcome Reward Models (ORMs), which rely solely on final answer supervision \cite{uesato2022solving,wang2024math,zhang2025lessonsdevelopingprocessreward}. By providing dense, step-by-step feedback, PRMs enable more reliable reasoning. Furthermore, they support advanced inference strategies like Tree-of-Thoughts (ToT) \cite{yao2023tree} and Monte Carlo Tree Search (MCTS) \cite{zhang2024rest}, where evaluating intermediate states is crucial. This success highlights a general principle: in specified generative tasks, verifying intermediate trajectories is essential for mitigating error accumulation, thereby ensuring robust and correct results.

We identify a natural correspondence between the chain-of-thought in reasoning and the coarse-to-fine generation trajectory of Semantic IDs. In the hierarchical ID structure, initial tokens serve as high-level semantic anchors (e.g., "Electronics"), while subsequent tokens refine the specific details (e.g., "Sony Headphones"). Analogous to a wrong initial step in a mathematical derivation, an error in a leading token acts as a spatial mis-routing, irreversibly diverting the generation into an incorrect semantic subspace. This renders all subsequent fine-grained predictions ineffective, as the model searches within a completely wrong item cluster. Current SID-based generative models lack the ability to evaluate intermediate steps, leaving them unable to detect and stop Semantic Drift before it becomes irreversible.

Motivated by these findings, we propose \textbf{\textsc{\modelname}} (\textbf{\underline{PRO}}cess Reward \textbf{\underline{M}}odels unlock Test-t\textbf{\underline{I}}me \textbf{\underline{S}}caling Laws in G\textbf{\underline{E}}nerative Recommendations), a novel framework that seamlessly integrates Process Reward Models into generative models. Our method comprises two core components: 

\begin{itemize}[leftmargin=*]
\item  A lightweight \textbf{path-level PRM}, trained end-to-end together with the generative backbone, is explicitly optimized to supervise intermediate inference steps. Crucially, unlike the generative backbone, which is trained via Teacher Forcing and thus blind to its own errors (Exposure Bias), the PRM is trained on sampled trajectories containing both positive and negative examples. This enables it to detect and penalize deviations in real-time. By arresting these deviations early, it effectively mitigates Semantic Drift, preventing the cascade of errors and ensuring recommendations remain aligned with user preferences.
\item A \textbf{PRM-guided Beam Search} strategy during inference. By leveraging dense feedback from the Path-level PRM, we prune low-quality branches early and explore high-potential semantic subspaces.
\end{itemize}Significantly, this approach unlocks \textbf{Test-Time Scaling Laws} for recommendation: by increasing the search width (inference compute), our smaller model outperforms larger models. This offers a flexible trade-off between latency and quality, allowing for latency-constrained search in industrial settings while maintaining lower computational costs compared to scaling model parameters.

Our main contributions can be summarized as follows:
\begin{itemize}[leftmargin=*]
\item We identify and formally define the Semantic Drift phenomenon in Generative Recommendation. We draw a novel parallel between the hierarchical generation of Semantic IDs and Chain-of-Thought reasoning in LLMs, highlighting the necessity of intermediate verification for robust retrieval.
\item We propose a novel framework that integrates PRMs into generative models. To our knowledge, this is the first work to employ dense supervision (Path-level PRM) and PRM-Guided Beam Search to align intermediate generation steps with user preferences.
\item We empirically demonstrate Test-Time Scaling Laws for Generative Recommendation. Our results show that scaling inference computing via PRM-guided search enables smaller models to outperform larger baselines, providing a flexible and efficient paradigm for industrial-scale recommendation.
\item We conduct extensive offline experiments and validate our method through online A/B tests on a large-scale platform, demonstrating significant improvements in core business metrics.
\end{itemize}

\section{Preliminary}

\subsection{Item Tokenization}
Each item $x\in\mathcal{X}$ is represented by an embedding $h \in \mathbb{R}^{d_h}$, which is quantized via $\mathcal{Q}: \mathbb{R}^{d_h} \rightarrow \{1, \dots, M\}^{d}$ into a $d$-layer discrete Semantic ID $[s_1, s_2, \dots, s_d]$ with codebook size $M$.

\subsection{Next-token Prediction}
For a user $u$ and context $c$, let $x_t$ be an interacted item. We extract $u$'s preceding interaction sequence $[x_1, ..., x_{t-1}]$. The goal of generation recommendations is to maximize the conditional probability of truly interacted item $x_t$ given $u$, $c$ and $u$'s historical sequence: $p_\theta\{x_t\ |\ x_1,\ ...,\ x_{t-1},\ u,\ c\}$. 
Item quantization allows generative models to predict $x_t$ in multiple steps.
For instance, target item $x_t$ is mapped into a Semantic ID path $[s_{t,1}, s_{t,2}, \dots, s_{t,d}]$. Therefore, it can be further transformed into maximizing the conditional probability from each inference step:
\begin{equation}\label{ntpsid}\small
    p_\theta(x_t\ |\ u,\ c) = \prod_{b=1}^dp_\theta\{s_{t,b}\ |\ s_{t,1},\ ...,\ s_{t,b - 1},\ x_1,\ ...,\ x_{t-1},\ u,\ c\}.
\end{equation}
With Eq. \ref{ntpsid}, NTP loss is defined as the negative log-likelihood of the ground-truth path:
\begin{equation}\label{eq:ntp_loss}\small
\mathcal{L}^{\text{NTP}}_{x_t} = - \sum_{b=1}^{d} \log p_\theta\left(s_{t,b} \mid s_{t,1}, \dots, s_{t,b-1}, x_1, \dots, x_{t-1}, u, c\right).
\end{equation}
During inference, beam search is employed to autoregressively generate the Semantic ID sequence step-by-step, where at each step the top-$K$ most probable paths are retained based on the conditional probabilities.

\subsection{Generative Backbone}
The encoder maps historical sequence $[x_1, \dots, x_{t-1}]$ to a hidden state sequence $\mathbf{E}^{(L)} \in \mathbb{R}^{(t-1) \times d_h}$. This is achieved through $L$ blocks of bidirectional self-attention and feed-forward networks.

The decoder takes the encoder output $\mathbf{E}^{(L)}$ and an autoregressively generated prefix of the target Semantic ID path $[s_{t,1}, \dots, s_{t,b-1}]$ to predict the next token $s_{t,b}$, using self-attention over the generated prefix, cross-attention over $\mathbf{E}^{(L)}$, and FFNs. Its final hidden state for the \(b\)-th position, denoted \(\mathbf{h}_{t,b}\), is projected via a linear layer and matched against a learnable codebook embedding matrix \(\mathbf{C}_b \in \mathbb{R}^{M \times d_h}\). The conditional probability for the next token is computed as a softmax over the dot product scores:
\begin{equation}\small
    p_\theta\left(s_{t,b} \mid s_{t,1}, \dots, s_{t,b-1}, x_1, \dots, x_{t-1}, u, c\right) = \frac{\exp\left(\mathbf{h}_{t,b}^\top \mathbf{C}_b[s]\right)}{\sum\limits_{m=1}^{M} \exp\left(\mathbf{h}_{t,b}^\top \mathbf{C}_b[m]\right)},
\end{equation}
where \(\mathbf{C}_b[s]\) denotes the embedding of token \(s\) in the \(b\)-th codebook. 

\subsection{Semantic Drift}\label{sd}
A critical challenge in Next-Token Prediction (NTP) for generative recommendation stems from the discrepancy between training and inference, formally known as exposure bias. During training, the model maximizes the likelihood of the next token conditioned on the ground-truth history (teacher forcing). This limits the model's exposure to its own prediction errors. Conversely, during inference, the model generates tokens autoregressively and relies on its previously generated sequence. Since the model is never exposed to erroneous intermediate states during training, early prediction errors inevitably accumulate.

In the specific context of Semantic IDs, we define this error propagation as \textbf{Semantic Drift}. Unlike unstructured text generation, Semantic IDs typically employ hierarchical quantization, where early tokens encode coarse-grained semantics (e.g., item categories) and later tokens capture fine-grained details. Consequently, a deviation in the initial steps causes the generation trajectory to diverge irreversibly into irrelevant semantic sub-spaces. Furthermore, when the model encounters these out-of-distribution (OOD) states, it tends to revert to the marginal distribution of the training data. This exacerbates popularity bias, leading the generated path toward generic, high-frequency items rather than specific, long-tail user preferences. To assess the extent of semantic drift, we later introduce a novel metric \(\text{HRecall}@b@k\) in Section \ref{hrecall} to quantify this phenomenon.

\section{Methodology}
\subsection{Overview}
\begin{figure*}[htbp]
    \centering
    \includegraphics[width=0.85\linewidth]{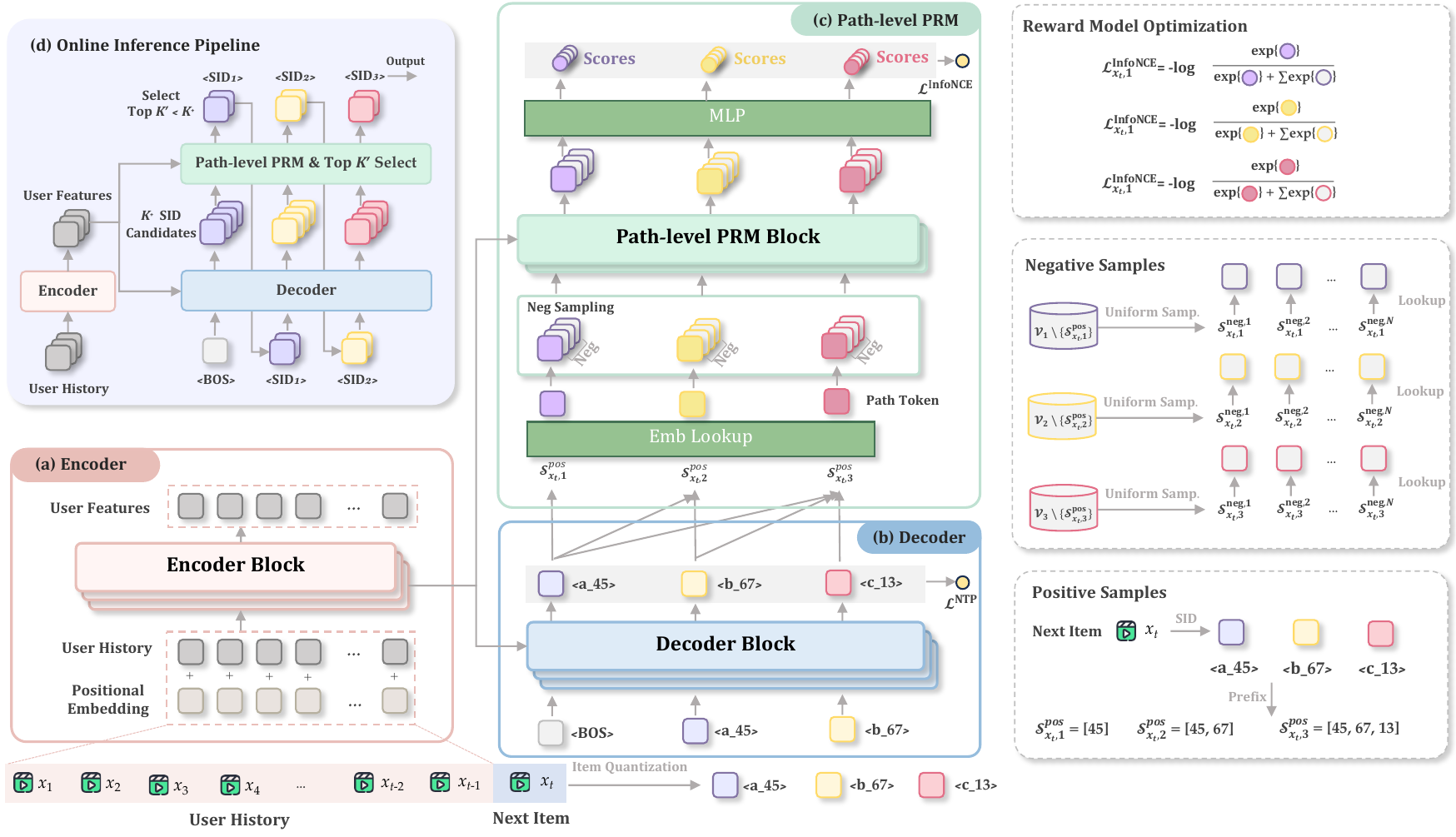}
    \caption{Overall framework of \textsc{\modelname}. (a) and (b): An encoder-decoder architecture optimized via the NTP loss. (c): The Path-level PRM assesses the quality of SID reasoning paths. During training, an InfoNCE loss is used to maximize the score of the ground-truth SID path relative to negative paths. (d): An illustration of Model inference. At each generation step, the encoder-decoder produces a candidate set of $K^+$ SIDs. These are then evaluated by the Path-level PRM, which selects the top $K'<K^+$ optimal tokens to proceed with the next step of autoregressive prediction.}
    \label{fig:ov} 
\end{figure*}

Fig. \ref{fig:ov} illustrates the overall framework of \textsc{\modelname}. 
We take an encoder-decoder approach, where the encoder takes user behavior sequence $[x_1, ..., x_{t-1}]$ as the input, and the decoder autoregressively generates the Semantic ID path. Upon that, Sec. \ref{prm} introduces a novel process reward model to address the problem of error propagation and accumulation in generative recommenders. 
Then, how process reward system achieves test-time scaling of generative models is explained in Sec. \ref{inference}.

\subsection{Path-Level PRM}\label{prm}
To address semantic drift, we propose a Process Reward Model (PRM) for generative recommendation. This section first introduces the definitions of positive and negative examples within the reward mechanism. We then describe the corresponding training  objective. Furthermore, although our reward mechanism is agnostic to the specific model architecture, to enable efficient online deployment, we propose a lightweight, low-latency attention-based model architecture suitable for industrial-scale recommendation scenarios, which will be detailed subsequently.

\subsubsection{Positive Samples. } The ground-truth item $x_t$ is quantized into a $d$-layer Semantic ID path \([s_{t,1}, s_{t,2}, \dots, s_{t,d}]\). Naturally, this complete path of length $d$ is the positive path for $x_t$ at depth $d$. Due to the nature of residual quantization, for the model to retrieve this positive path at depth $d$, it must be able to retrieve the corresponding path \([s_{t,1}, s_{t,2}, \dots, s_{t,d-1}]\) at depth $d-1$. 
Hence, the positive path at depth  $b$ is its prefix of length $b$:
\begin{equation}
    \mathcal{S}_{x_t,b}^{\text{pos}}=[s_{t,1},\ \dots,\ s_{t,b}],\quad b \leq d,
\end{equation}
where $b$ denotes the path length.

\subsubsection{Negative Sampling Strategy.} The negative sampling strategy within our reward mechanism is key to mitigating semantic drift. Unlike NTP, which learns solely from positive examples, our proposed PRM introduces a negative sampling strategy.
By being trained on both positive and negative samples, the PRM acquires the ability to discriminate between relevant and irrelevant Semantic ID paths. 
This capacity allows it to identify erroneous intermediate SID reasoning paths during inference, thereby alleviating the semantic drift problem.
Empirical experiments in Sec. \ref{sec:as} illustrate that by eliminating  prediction errors, recommendation results can be more aligned with user preferences,  leading to improved recommendation metrics.

During the construction of the quantization model, the allocated codebook space often far exceeds the number of actual items. As a result, not all Semantic ID combinations correspond to real items.  We refer to paths that map to existing items as \textit{valid paths}. Let \(\mathcal{V}_b\) denote the set of all \textit{valid} Semantic ID paths of length \(b\) that correspond to existing items in the catalog. For a given positive item \(x_t\) with its positive path prefix \(\mathcal{S}_{x_t,b}^{\text{pos}} \in \mathcal{V}_b\) at depth \(b\), we construct a negative sample set \(\mathcal{N}_{x_t, b}\) by uniformly sampling \(N\) distinct paths from the valid path set, excluding the positive path itself:
\begin{equation}\small
    \mathcal{N}_{x_t, b} = \{ \mathcal{S}^{\text{neg}, i} \mid \mathcal{S}^{\text{neg}, i} \sim \text{Uniform}(\mathcal{V}_b \setminus \{\mathcal{S}_{x_t,b}^{\text{pos}}\}),\ i = 1, \dots, N \}.
\end{equation}
Thus, the PRM is exposed not only to relevant user-SID combinations but also to irrelevant ones. The inclusion of these negative training samples enables the PRM to recognize erroneous intermediate states during inference.

\subsubsection{Reward Models Optimization.}
Based on the above definitions, the PRM learns a mapping \(\mathcal{F}\): given a user \(u\), context \(c\), and a candidate SID sequence of length \(b\) (where \(1 \leq b \leq d\)), it outputs a relevance score for that sequence with respect to the user:
\begin{equation}
    \mathcal{F}: (u, c, [s_1, \dots, s_b]) \mapsto y \in \mathbb{R}.
\end{equation}
Note that the proposed process reward mechanism is fundamentally agnostic to the specific choice of model architecture implementing \(\mathcal{F}\).

For each positive sample $x_t$, there exists a corresponding positive path across the $d$ Semantic ID layers. Considering the $b$-th layer, where $1 \leq b \leq d$, the positive path $\mathcal{S}_{x_t,b}^{\text{pos}}=[s_{t,1},\ \dots,\ s_{t,b}]$ is fed into the $\mathcal{F}$ module in parallel with $N$ sampled negative paths. The positive logit \(y_{\mathcal{S}_{x_t,b}^{\text{pos}}}\) is obtained by \(\mathcal{F}(u, c, \mathcal{S}_{x_t,b}^{\text{pos}})\). For each negative path \(\mathcal{S}^{\text{neg}} \in \mathcal{N}_{x_t, b}\), we compute its logit \(y_{\mathcal{S}^{\text{neg}}} = \mathcal{F}(u, c, \mathcal{S}^{\text{neg}})\).

The InfoNCE loss is employed at depth \(b\) to maximize the logit for the positive path relative to the sampled negatives:
\begin{equation}
    \mathcal{L}^{\text{InfoNCE}}_{x_t,b}=-\log\frac{\exp\{y_{\mathcal{S}_{x_t,b}^{\text{pos}}}\}}{\exp\{y_{\mathcal{S}_{x_t,b}^{\text{pos}}}\} + \sum_{\mathcal{S}^{\text{neg}} \in \mathcal{N}_{x_t, b}} \exp\{y_{\mathcal{S}^{\text{neg}}}\}}.
\end{equation}

We sum the InfoNCE losses across different path-levels:
\begin{equation}
\mathcal{L}^{\text{InfoNCE}}_{x_t}=\sum_{b=1}^d\mathcal{L}^{\text{InfoNCE}}_{x_t,b}.
\end{equation}
The next-token prediction task and the PRM task are trained jointly in an end-to-end manner. The complete loss function is defined as:
\begin{equation}
\mathcal{L}^{\text{Total}}_{x_t}=\mathcal{L}^{\text{NTP}}_{x_t}+\mathcal{L}^{\text{InfoNCE}}_{x_t}.
\end{equation}

\subsubsection{Lightweight PRM Architecture.} While the mapping function  $\mathcal{F}: (u, c, [s_1, \dots, s_b]) \mapsto y \in \mathbb{R}$ is architecture-agnostic, we propose a specific lightweight, low-latency PRM design for practical deployment. This design utilizes cross-attention for GPU-parallelizable acceleration, enabling rapid scoring of a large set of candidate paths. Furthermore, it reuses encoder-side features, leveraging the encoder's powerful representation capacity to reduce the parameter count of the PRM module.

For any SID depth \(b \in \{1, \dots, d\}\) and SID path \(\mathcal{S}=[s_{\cdot,1}, \dots, s_{\cdot,b}] \in \{1,\ ...,\ M\}^b\), we define \(P_{\mathcal{S}}\) as the representation vector corresponding to this path. Each path is mapped to a unique embedding vector, which serves as the target-side input for the path-level PRM:
\begin{equation}\label{dis1}
    P_{\mathcal{S}} = \text{EmbLookup}([s_{\cdot,1},\ \dots,\ s_{\cdot,b}]).
\end{equation}
Note that we do not use the raw SID tokens for representation. Instead, to ensure efficiency, we represent them with a single token in order to reduce attention computation.

We employ a path-level cross-attention mechanism to evaluate any intermediate path. 
Specifically, the intermediate path representation \(P_{\mathcal{S}}\) serves as the query.
The final encoder output \(E^{(L)}\), which captures rich user interest information through multiple self-attention layers over the historical sequence—is reused as the key and value in this component, ensuring efficient feature reuse alongside the main generation task.

We stack $F$ layers of cross-attention with residual connections, each followed by a feed-forward network. For layer   \(i \in \{1,...,F\}\):
\begin{equation}\label{dis2}\begin{aligned}
    P_{\mathcal{S}}^{(i)'} &= P_{\mathcal{S}}^{(i)}+\text{CrossAttn}(P_{\mathcal{S}}^{(i-1)},\ E^{(L)},\ E^{(L)}),\\
    P_{\mathcal{S}}^{(i)} &= P_{\mathcal{S}}^{(i)'} + \text{FFN}(\text{RMSNorm}(P_{\mathcal{S}}^{(i)'})).
\end{aligned}\end{equation}

After processing through multiple stacked layers of path-level cross-attention, the candidate path representation has been sufficiently interacted with the user representation. A subsequent Multi-Layer Perceptron (MLP) then outputs a single scalar logit. This logit serves as the reward score for the path \(\mathcal{S}=[s_{\cdot,1}, \dots, s_{\cdot,b}]\), conditioned on \(u\) and \(c\):
\begin{equation}\label{dis3}
    y_{\mathcal{S}} = MLP(P_{\mathcal{S}}^{(F)}).
\end{equation}

\subsection{Inference Strategy}\label{inference}

\begin{algorithm}
\caption{Beam Search with Path-level PRM}
\label{alg:prm_beam}
\begin{algorithmic}[1]
\STATE Initialize $\mathcal{B}^{(0)} \leftarrow \{[\text{BOS}]\}$ \hfill\COMMENT{Initial beam with BOS token}
\STATE $\mathcal{S}^{(0)} \leftarrow \{0\}$ \hfill\COMMENT{Initial scores}
\STATE Compute encoder output $E^{(L)} = E([x_1,...,x_{t-1}])$ 
\FOR{$b = 1$ \TO $d$}
    \STATE $\mathcal{C} \leftarrow \emptyset$ \hfill\COMMENT{Candidate paths}
    \STATE $\mathcal{S}_C \leftarrow \emptyset$ \hfill\COMMENT{Candidate scores}
    \FOR{each path $\mathcal{P} \in \mathcal{B}^{(b-1)}$ with score $s \in \mathcal{S}^{(b-1)}$}
        \STATE Compute decoder output $D^{(L)}$ for $\mathcal{P}$
        \STATE Compute distribution $p(s_b) = \text{softmax}({D^{(L)}_b}^T C_b)$
        \STATE Extract top-$K$ tokens $\{(s_b^{(k)}, p_k)\}_{k=1}^K$ where $p_k = p(s_b^{(k)})$
        \FOR{$k = 1$ \TO $K$}
            \STATE $\mathcal{P}' \leftarrow \mathcal{P} \oplus s_b^{(k)}$ \hfill\COMMENT{Path extension}
            \STATE $s' \leftarrow s + \log p_k$ \hfill\COMMENT{Score update}
            \STATE $\mathcal{C} \leftarrow \mathcal{C} \cup \{\mathcal{P}'\}$
            \STATE $\mathcal{S}_C \leftarrow \mathcal{S}_C \cup \{s'\}$
        \ENDFOR
    \ENDFOR
    \STATE Select top-$K^+$ paths candidates $\mathcal{C}_{K^+}$ from $\mathcal{C}$ based on $\mathcal{S}_C$
    \STATE Use path-level PRM Eq. (\ref{dis1} - \ref{dis3}) to calculate path scores $\mathcal{Y}_{\mathcal{C}_{K^+}}$ on paths candidates $\mathcal{C}_{K^+}$
    \STATE Select top-$K'$ paths candidates $\mathcal{C}_{K'}$ from $\mathcal{C}_{K^+}$ based on $\mathcal{Y}_{\mathcal{C}_{K^+}}$  \hfill\COMMENT{Rank intermediate path qualities}
    \STATE Update $\mathcal{B}^{(b)} \leftarrow \{\mathcal{P}_{(1)},...,\mathcal{P}_{(K')}\}$
    \STATE Update $\mathcal{S}^{(b)} \leftarrow \{s_{(1)},...,s_{(K')}\}$
\ENDFOR
\RETURN $\mathcal{B}^{(d)}$ \hfill\COMMENT{Top-$K'$ complete paths of length $d$}
\end{algorithmic}
\end{algorithm}

\begin{figure}[htbp]
    \centering
    \includegraphics[width=\linewidth]{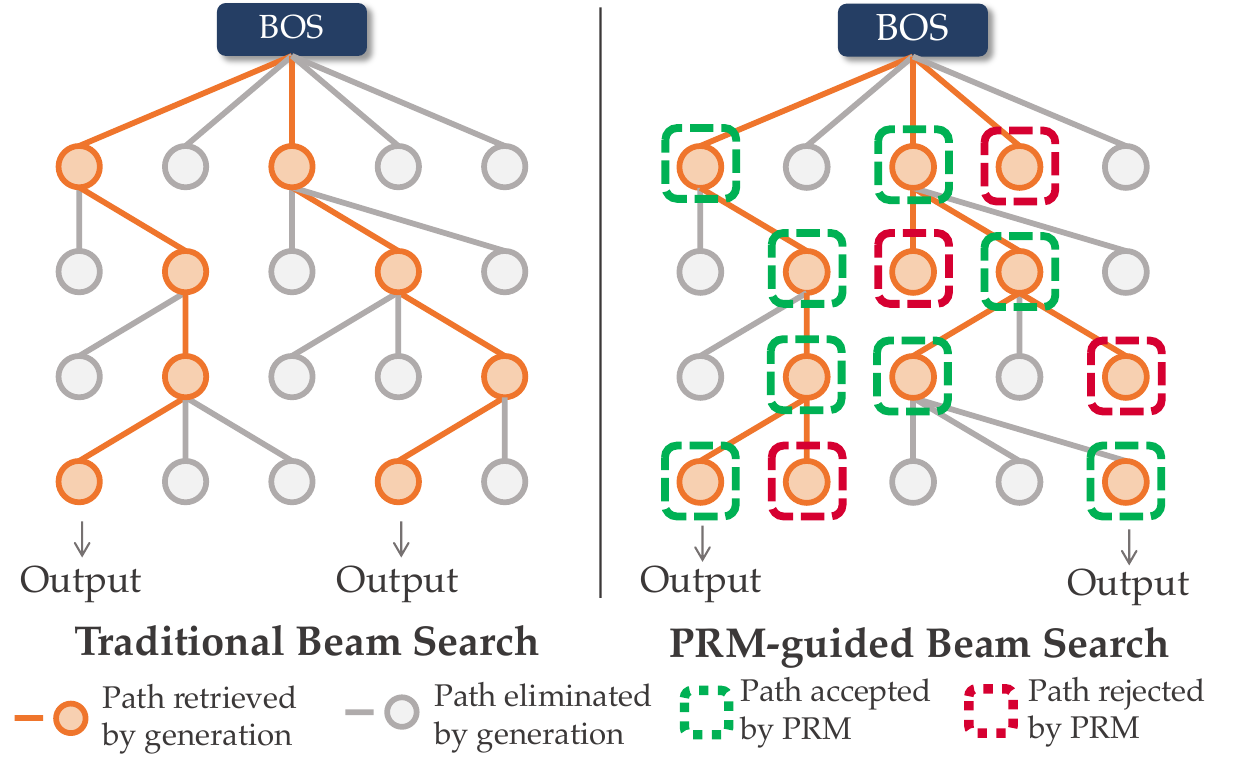}
    \caption{Illustration of our test-time scaled beam search method compared with traditional beam search for SID generation.}
    \label{fig:beamsearch} 
\end{figure}
Pioneering research in LLMs has demonstrated that by reasonably increasing test-time computational resources, smaller models can surpass the capabilities of larger ones. Similarly, after constructing our process reward model, we can enhance the capability of the generative recommendation model by increasing the number of candidates considered at test time.

Specifically, in generative recommenders without PRMs, standard beam search is typically employed with a fixed beam size, denoted as $K$. After computing the dot product between the codebook and the decoder's output vector, the top-$K$ most promising paths are selected for the next decoding step. Simply increasing $K$ would allow the model to consider more SID path possibilities, but it would also linearly increase the computational complexity of each decoder step, making it an impractical test-time scaling strategy.

Our designed lightweight path-level PRM can efficiently and in parallel evaluate the scores of intermediate reasoning trajectories. 
As shown in Algorithm. \ref{alg:prm_beam}, when the beam size is increased to \(K^+ \gg K\), the path-level PRM can serve as a process reward model to score these paths and select a much smaller subset \(K' \ll K^+\) in descending order of scores to feed into the subsequent decoding step.
In particular, if we set \(K'\) to match the original beam size (i.e., \(K' = K\)), the computational resources on the decoder side remain unchanged before and after scaling. The only additional cost comes from the lightweight cross-attention computation. This allows for a significant improvement in the quality of generation at each step, and consequently in the final recommendation accuracy, with only a marginal increase in inference cost.

In Section \ref{exp}, we provide a detailed analysis of the effectiveness of this test-time scaling scheme. Subsequent empirical experiments verify that the additional computation introduced by the path-level cross attention is negligible compared to the entire generative decoding process, ensuring efficient inference-time scaling.

\section{Experiments}\label{exp}
We conduct extensive empirical experiments on both  mass real industrial data and public benchmarks to answer the following research questions:\\
\textbf{RQ1}: How well does our proposed generative recommender with process reward mechanism outperform strong  discriminative and generative recommendation baselines?\\
\textbf{RQ2}: Can online A/B tests in a real industrial environment demonstrate positive user feedback from our method?\\
\textbf{RQ3}: How does the process reward mechanism help mitigate semantic drift?\\
\textbf{RQ4}: How does PRM contribute to overall performance at each step?\\
\textbf{RQ5}: Are the test-time scaling laws in generative recommenders validated?\\
\textbf{RQ6}: Is test-time scaling via the path-level PRM more efficient than simply scaling up model parameters?\\
\textbf{RQ7}: How do hyper-parameter settings affect the model performance?\\
\vspace{-0.5cm}

\subsection{Offline Public Dataset Experiments (RQ1)}
\begin{table*}[!ht]
    \caption{Performance of \textsc{\modelname} compared with baselines on public datasets. \textbf{Bold} indicates the best result, \underline{underline} indicates the second best. Baseline results are taken from the original papers or reported under identical settings \cite{rajput2023recommender,houactionpiece}. Experiments show that the proposed method outperforms baselines in Recall@\textit{k} and NDCG@\textit{k} ($k\in\{5,10\}$).}
    \centering
    \label{tab:overall}
\resizebox{1.\linewidth}{!}{\begin{tabular}{clccccccccc}
        \toprule
\multicolumn{2}{c}{\multirow{2}{*}{\textbf{Methods}}} & \multicolumn{4}{c}{\textbf{Sports and Outdoors}}                              & \multicolumn{4}{c}{\textbf{Beauty}}                                            \\
            \cmidrule(lr){3-6} \cmidrule(lr){7-10}
           & & \textbf{Recall@5}           & \textbf{NDCG@5}          & \textbf{Recall@10}           & \textbf{NDCG@10}         & \textbf{Recall@5}            & \textbf{NDCG@5}          & \textbf{Recall@10}           & \textbf{NDCG@10}         \\\midrule
\multirow{6}{*}{\textbf{Traditional}} &GRU4REC \cite{jannach2017recurrent}    & 0.0129         & 0.0086          & 0.0204          & 0.0110          & 0.0164          & 0.0099          & 0.0283          & 0.0137          \\
&Caser \cite{tang2018personalized}      & 0.0116         & 0.0072          & 0.0194          & 0.0097          & 0.0205          & 0.0131          & 0.0347          & 0.0176          \\
&HGN \cite{ma2019hierarchical}        & 0.0189         & 0.0120          & 0.0313          & 0.0159          & 0.0325          & 0.0206          & 0.0512          & 0.0266          \\
&S$^3$-Rec \cite{zhou2020s3}     & 0.0251         & 0.0161          & 0.0385          & 0.0204          & 0.0387          & 0.0244          & 0.0647          & 0.0327          \\
&Bert4Rec \cite{sun2019bert4rec}   & 0.0115         & 0.0075          & 0.0191          & 0.0099          & 0.0203          & 0.0124          & 0.0347          & 0.0170          \\
&SASRec \cite{kang2018self}     & 0.0233         & 0.0154          & 0.0350          & 0.0192          & 0.0387          & 0.0249          & 0.0605          & 0.0318          \\ \midrule
\multirow{4}{*}{\textbf{Generative}} &TIGER \cite{rajput2023recommender}      & 0.0264         & 0.0181          & 0.0400          & 0.0225          & 0.0454          & 0.0321          & 0.0648          & 0.0384          \\
&HSTU \cite{zhai2024actions}       & 0.0258         & 0.0165          & 0.0414          & 0.0215          & 0.0469          & 0.0314          & 0.0704          & 0.0389          \\
&ActionPiece \cite{houactionpiece} & \underline{0.0316}   & \underline{0.0205}    & \underline{0.0500}      & \underline{0.0264}    & \underline{0.0511}    & \underline{0.0340}     & \underline{0.0775}    & \underline{0.0424}    \\ 
&\textbf{\modelname}  & \textbf{0.0450} & \textbf{0.0296} & \textbf{0.0689} & \textbf{0.0373} & \textbf{0.0536} & \textbf{0.0345} & \textbf{0.0821} & \textbf{0.0437}\\ \midrule
& \textbf{Improv.} & \textbf{+42.41\%} & \textbf{+44.39\%} & \textbf{+37.80\%} & \textbf{+42.19\%} & \textbf{+4.90\%} & \textbf{+1.47\%} & \textbf{+5.60\%} & \textbf{+3.07\%} \\
\bottomrule
    \end{tabular}}
\end{table*}

\subsubsection{Datasets.} The Amazon Review dataset ~\cite{mcauley2015image} is commonly used as a benchmark for generative recommenders. It contains user reviews and corresponding item metadata from Amazon spanning from May 1996 to July 2014. The dataset includes multiple categories; we select the \textbf{Beauty} and \textbf{Sports and Outdoors} categories for experiments. For data preprocessing, as well as training, validation, and test set splitting, we follow the same protocol as TIGER ~\cite{rajput2023recommender}. Users with fewer than five reviews are filtered out. User interaction sequences are sorted chronologically; the last item in each sequence is held out for testing, the second-to-last item is used for validation, and the remaining items are used for model training.

\subsubsection{Baselines.} We select baselines from two perspectives. First, traditional sequential modeling methods include:
\begin{itemize}[leftmargin=*]
\item \textbf{GRU4REC} \cite{jannach2017recurrent}: Classic sequential modeling approach based on RNNs.
\item \textbf{Caser} \cite{tang2018personalized}: A sequential recommendation model that employs convolutional operations to capture user preferences.
\item \textbf{HGN} \cite{ma2019hierarchical}: Enhancing traditional sequential recommendation through an efficient hierarchical gating mechanism.
\item \textbf{S$^\mathbf{3}$-Rec} \cite{zhou2020s3}: Improving sequential recommendation capability by using self-supervised pretraining to enrich feature representations.
\item \textbf{Bert4Rec} \cite{sun2019bert4rec}: A model that employs bidirectional self-attention to model user behavior sequences.
    \item \textbf{SASRec} \cite{kang2018self}: Causal self-attentions for next-item predictions with binary cross-entropy loss.
\end{itemize}
Secondly, generative recommendation methods include:
\begin{itemize}[leftmargin=*]
    \item \textbf{TIGER} \cite{rajput2023recommender}: TIGER adopts a Transformer-based sequence-to-sequence approach, in which the decoder autoregressively predicts Semantic IDs of the target item.
    \item \textbf{HSTU} \cite{zhai2024actions}: A model for observing scaling laws in industrial recommendation scenarios.
    \item \textbf{ActionPiece} \cite{houactionpiece}: ActionPiece explicitly incorporates contextual information when tokenizing action sequences.
\end{itemize}

\subsubsection{Implementation details.} We employ RQ-VAE to quantize items into a 3-level codebook with size \(M = 256\). The model is trained using the Adam optimizer with a learning rate of 0.001 for 200 epochs. Early stopping is applied with a patience of 20. The checkpoint achieving the best validation performance is used for testing. The input sequence length for the encoder is set to 40. The hidden size is 256. 

\subsubsection{Metrics.} Following \citet{rajput2023recommender}, we employ $\text{Recall}@k$ and $\text{NDCG}@k$ for evaluation, specifically at $k = 5$ and $k = 10$.

\subsubsection{Results.}  Experimental results are illustrated in Table. \ref{tab:overall}. 
Traditional sequential modeling methods perform relatively weaker than generative approaches. Among generative recommendation methods, Semantic‑ID‑based approaches, owing to their coarse‑to‑fine generation scheme, achieve better results. 

Compared with these strong Semantic‑ID‑based baselines, \textsc{\modelname}, through its process‑reward mechanism and test‑time scaling strategy, amplifies the benefits of Semantic‑ID‑based generative recommendation. After effectively addressing semantic drift, our model demonstrates state‑of‑the‑art performance. Against the best baseline, ActionPiece~\cite{houactionpiece}, \textsc{\modelname} shows superior results across different datasets and metrics. This improvement is attributed to the proposed path‑level PRM, which mitigates the semantic‑drift issue overlooked in conventional beam search, reduces error accumulation during search, and thereby leads to promising experimental outcomes.
\begin{table*}[!ht]
    \caption{Overall performance on industrial-scale dataset. The best results are in \textbf{bold}. The second-best are \underline{underlined}. Our method significantly outperforms all baselines (paired t-test, $p < 0.05$).}
    \centering
    \label{tab:industrial_results}
    \begin{tabular}{lcccccc}
        \toprule
        \textbf{Method} & \textbf{Recall@100} & \textbf{NDCG@100} & \textbf{Recall@500} & \textbf{NDCG@500} & \textbf{Recall@1000} & \textbf{NDCG@1000} \\
        \midrule
        GRank \cite{sun2025grank} & \underline{0.1010} & \underline{0.00472} & \underline{0.2396} & \underline{0.01089} & \underline{0.3178} & \underline{0.01422} \\
        MPFormer \cite{sun2025mpformer} & 0.0706 & 0.00331 & 0.1581 & 0.00694 & 0.2010 & 0.00843 \\
        MISS \cite{guo2025miss} & 0.0844 & 0.00283 & 0.2033 & 0.00741 & 0.2441 & 0.00892 \\
        GPRP \cite{zheng2024full} & 0.0414 & 0.00147 & 0.0661 & 0.00299 & 0.0929 & 0.00415 \\
        Kuaiformer \cite{liu2024kuaiformer} & 0.0622 & 0.00293 & 0.1388 & 0.00625 & 0.1761 & 0.00796 \\
        CRM \cite{liu2024crm} & 0.0409 & 0.00215 & 0.0977 & 0.00485 & 0.1437 & 0.00683 \\
        \midrule
        \textbf{\modelname\ {\small ($K^+=4000$)}} & 
        \textbf{0.1494} {\scriptsize (+47.92\%)} & 
        \textbf{0.00652} {\scriptsize (+38.14\%)} & 
        \textbf{0.2836} {\scriptsize (+18.36\%)} & 
        \textbf{0.01231} {\scriptsize (+13.04\%)} & 
        \textbf{0.3358} {\scriptsize (+5.66\%)} & 
        \textbf{0.01445} {\scriptsize (+1.62\%)} \\
        
        \textbf{\modelname\ {\small ($K^+=6000$)}} & 
        \textbf{0.1609} {\scriptsize (+59.31\%)} & 
        \textbf{0.00663} {\scriptsize (+40.47\%)} & 
        \textbf{0.3017} {\scriptsize (+25.92\%)} & 
        \textbf{0.01272} {\scriptsize (+16.80\%)} & 
        \textbf{0.3637} {\scriptsize (+14.44\%)} & 
        \textbf{0.01504} {\scriptsize (+5.77\%)} \\
        \bottomrule
    \end{tabular}
\end{table*}
\subsection{Industrial-scale Experiments (RQ1)}
\subsubsection{Datasets.} Although our method shows improvements on public datasets, real-world industrial environments involve significantly larger data scales, longer user sequences, and more severe exposure bias, meaning that gains on public benchmarks may not fully reflect practical value. Therefore, we conduct additional performance evaluations on Kuaishou—one of the world’s largest short-video platforms.

We train our model (including baselines) via online learning using real user logs from the Kuaishou app. These logs cover over 400 million daily active users, generate approximately 50 billion user interactions per day, and involve more than 100 million items optimized daily by the model. To handle such a large item corpus, we quantize their multi-modal representations using Residual K-means \cite{luo2025qarm} with a tokenizer of depth \(d=3\) and codebook size \(M=8192\).

Furthermore, we will later present results from online A/B tests conducted on real users to further validate the practical benefits of \textsc{\modelname} in industrial settings.

\subsubsection{Baselines.} We compare the proposed method with six publicly published recommendation models that have been fully deployed on the industrial environments:
\begin{itemize}[leftmargin=*]
    \item \textbf{GRank} \cite{sun2025grank}: A structured-index-free retrieval paradigm with an end-to-end ranking module.
    \item \textbf{MPFormer} \cite{sun2025mpformer}: A transformer-based recommendation model employing multi-objective estimation.
    \item \textbf{MISS} \cite{guo2025miss}: A multi-modal tree-based retrieval integrating both collaborative and multi-modal sequence search.
    \item \textbf{GPRP} \cite{zheng2024full}: A model that addresses selection biases in cascaded recommendation systems.
    \item \textbf{Kuaiformer} \cite{liu2024kuaiformer}: A transformer-based approach using next-token prediction and Approximate Nearest Neighbor search.
    \item \textbf{CRM} \cite{liu2024crm}: A transformer-based retrieval model with controllable conditioning.
\end{itemize}

\subsubsection{Implementation Details.} The hidden size is set to 1024. In PRM-guided beam search, the global beam size \(K\) is set equal to \(K'\) (\(K = K' = 1000\)). Regarding the expanded candidate size \(K^+\), we report results for both 4000 and 6000. The user sequence length is 256. The numbers of encoder and decoder blocks are set to 4, while the number of PRM blocks is set to 1 to ensure inference efficiency.

\subsubsection{Metrics.} We use \(\text{Recall}@k\) and \(\text{NDCG}@k\) to evaluate. Since models are often required to return hundreds to thousands of items in online production environments, we set \(k \in \{100, 500, 1000\}\).

\subsubsection{Results.}  Experimental results are reported in Tab. \ref{tab:industrial_results}. From these results, we observe that:
\begin{itemize}[leftmargin=*]
    \item \textbf{\textsc{\modelname} consistently outperforms all baselines.}  Compared to the strongest baseline, relative improvements reach \textbf{47.92\%} and \textbf{59.31\%} in $\text{Recall}@100$ for $K^+=4000$ and $K^+=6000$, respectively. This indicates that the proposed path-level PRM, serving as a process reward model, can more accurately capture user preferences and improve recommendation metrics.
    \item \textbf{\textsc{\modelname} exhibits greater advantages at smaller retrieval sizes.}  In particular, the improvement in \(\text{NDCG}@100\) over the strongest baseline is as high as \textbf{40.47\%}. This shows that \textsc{\modelname} effectively ranks items of interest higher in the generated result set, demonstrating the strong discriminative capability of the path-level PRM.
\end{itemize}

\subsection{Online A/B Test Result (RQ2)}
\begin{table*}[!ht]
    \caption{Online A/B Test Results of \textsc{\modelname}. Confidence intervals (CI) are calculated with 0.05 significance level.}
    \centering
    \label{tab:ab}
    \resizebox{.8\linewidth}{!}{\begin{tabular}{lcccccc}
            \toprule
    \textbf{Apps}          & \textbf{\makecell{Total App\\ Usage Time}} & \textbf{\makecell{Total App\\ Usage Time (CI)}} & \textbf{\makecell{App Usage Time\\ per User}} & \textbf{\makecell{App Usage Time\\ Per User (CI)}} & \textbf{\makecell{Total Video\\ Watch Time}} & \textbf{\makecell{Watch Time\\ per Video View}} \\
     \midrule
    Kuaishou      & +0.121\%                                                       & {[}+0.04\%, +0.20\%{]}                                              & +0.120\%                                                          & {[}+0.06\%, +0.18\%{]}                                                 & +0.431\%                                                   & +0.440\%                                                            \\
    Kuaishou Lite & +0.131\%                                                       & {[}+0.03\%,+ 0.23\%{]}                                              & +0.160\%                                                          & {[}+0.07\%,+ 0.25\%{]}                                                 & +0.398\%                                                   & +0.296\%                                                            \\                                                                \bottomrule
    \end{tabular}}
\end{table*}

To validate the practical value of the proposed method in a real industrial setting, we conducted online A/B tests on two short-video applications: Kuaishou and Kuaishou Lite. We allocate 5\% of the total users (around 20 million users) to the experimental group and another 5\% to the control group for each app. The control group used a generative recommendation model with a traditional beam search, while the experimental group applied the PRM with a PRM-guided search. Detailed configurations are provided in Section \ref{sec:costs}. The experiment lasted for 7 days.

Results are presented in Table \ref{tab:ab}, showing that \textsc{\modelname} significantly increases the time users spend watching videos within the apps. At a 95\% confidence level, users in the experimental group exhibited significantly higher app usage time compared to the baseline. Specifically, on Kuaishou Lite, we observed an increase in \textbf{total app usage time by 0.131\%} and in \textbf{app usage time per user by 0.160\%}. These results strongly demonstrate the practical value of our method in industrial recommendation scenarios.

\subsection{Ablation Studies (RQ3 \& RQ4)}\label{sec:as}
\begin{table}[!ht]
    \caption{Ablation experiment results on industrial dataset. The best results are highlighted in \textbf{bold}. \(K^+_b\) denotes the size of the candidate set scored by the PRM at the \(b\)-th Semantic ID generation step. If no PRM is applied, it is marked as "-"; otherwise, the candidate set size is set to \(K^+_b = 4000\).}
    \centering
    \label{tab:prm_performance}
    \resizebox{\linewidth}{!}{\begin{tabular}{c@{\hspace{0.5em}}cccccc}
        \toprule
        \multirow{2}{*}{\textbf{ID}} & \multicolumn{3}{c}{\textbf{PRM Settings}} & \multicolumn{3}{c}{\textbf{Performance}} \\
        \cmidrule(lr){2-4} \cmidrule(lr){5-7}
        & \textbf{\textit{K}}$\mathbf{^+_1}$ & \textbf{\textit{K}}$\mathbf{^+_2}$ & \textbf{\textit{K}}$\mathbf{^+_3}$ & \textbf{\makecell{HRecall@1\\@1000}} & \textbf{\makecell{HRecall@2\\@1000}} & \textbf{\makecell{HRecall@3\\@1000}} \\
        \midrule
        \small{(1)} & - & - & - & 0.9238 & 0.3718 & 0.2296 \\ 
        \small{(2)} & 4000 & - & - & \textbf{0.9431} & 0.3738 & 0.2322 \\ 
        \small{(3)} & - & 4000 & - & 0.9238 & 0.4435 & 0.2544 \\ 
        \small{(4)} & - & - & 4000 & 0.9238 & 0.3718 & 0.2650 \\ 
        \small{(5)} & - & 4000 & 4000 & 0.9238 & 0.4435 & 0.3199 \\ 
        \small{(6)} & 4000 & - & 4000 & \textbf{0.9431} & 0.3738 & 0.2746 \\ 
        \small{(7)} & 4000 & 4000 & - & \textbf{0.9431} & \textbf{0.4711} & 0.2633 \\ 
        \small{(8)} & 4000 & 4000 & 4000 & \textbf{0.9431} & \textbf{0.4711} & \textbf{0.3358} \\
        \bottomrule
    \end{tabular}}
\end{table}
We conduct ablation experiments on an industrial dataset, as presented in Table \ref{tab:prm_performance}. The global beam size is fixed at \(K = 1000\) across all settings. Our experiments comprehensively explore all possible configurations: applying no discrimination at any step, applying it at only one step, at two steps, and at all three steps.

\subsubsection{Metrics.}\label{hrecall} We propose \textbf{Hierarchical Recall} (\(\text{HRecall}@b@k\)) to measure semantic drift by evaluating the rate at which recommendation performance degrades as the intermediate reasoning step progresses. This definition is generalized from \citet{guo2025miss}:  
Assume the beam search runs for \(d\) steps, and let \(\mathcal{S}\) be the set of ground-truth items. Define \(\mathcal{S}^{(b)} = \{ s[:b] \mid s \in \mathcal{S} \}\) as the set of ground-truth prefixes at step \(b\) (where \(b \leq d\)). Let \(\mathcal{B}^{(b)}\) be the set of $k$ paths retained by the beam at step \(b\). The hierarchical recall at step \(b\) is defined as:
\begin{equation}
    \text{HRecall}@b@k=\frac{|\mathcal{B}^{(b)}\bigcap\mathcal{S}^{(b)}|}{|\mathcal{S}^{(b)}|}.
\end{equation}
Based on this metric, we can fairly compare the generation quality at intermediate steps across different configurations. When \(b = d\), the hierarchical recall is equivalent to the standard recall metric, i.e., $\text{HRecall}@d@k = \text{Recall}@k$.

\subsubsection{Results.} Overall, the results in Table \ref{tab:prm_performance} confirm that applying the PRM at each step improves performance.  

For RQ3, which examines whether the proposed method alleviates semantic drift by filtering erroneous intermediate reasoning steps, we analyze the comparisons between configurations. Comparing (1) vs. (2)–(4) reveals that reducing intermediate reasoning error at a single step leads to improved overall recommendation performance. Similarly, results for (5)–(7), where the PRM is active at two steps, show further gains compared to activating it at only one step. (8), where the PRM is applied at all three reasoning steps, continues to improve over (5)–(7). These results demonstrate that error accumulation in Semantic ID generation is effectively mitigated, confirming that the semantic drift can be resolved.

Regarding RQ4, which aims to verify whether the proposed PRM contributes to overall performance at each step of multi-step Semantic ID reasoning, the ablation study confirms that our designed process reward mechanism is effective at every reasoning step. Its cumulative application leads to consistent improvements in recommendation performance.

\begin{figure*}[htbp]
    \centering
    \includegraphics[width=\linewidth]{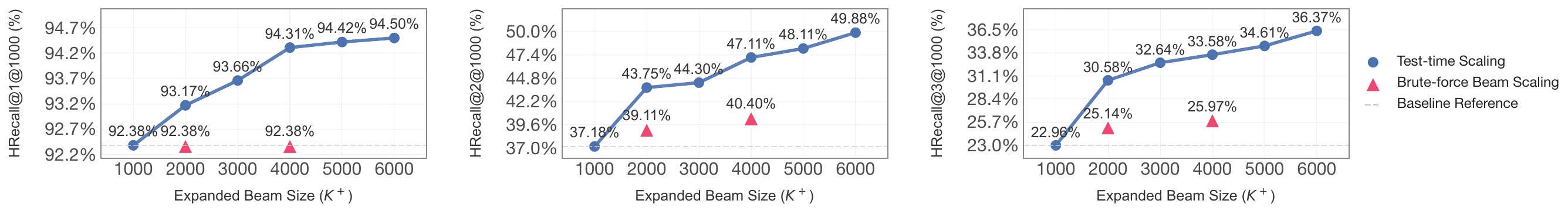}
    \caption{Test-time scaling laws can be validated by expanding \(K^+\) in \textsc{\modelname}. The red annotations indicate results of brute-force increasing the global beam size \(K\), which significantly increases decoder computation. In contrast, our lightweight path-level PRM enables significant metric improvements through test-time scaling without adding decoder computation.}
    \label{fig:scale}
\end{figure*}

\subsection{Validation of Test-time Scaling Laws (RQ5)}
To validate the test-time scaling laws in generative recommendation systems—namely, that increasing computation during inference improves recommendation performance, we progressively increase the number of path candidates \(K^+\) following Algorithm \ref{alg:prm_beam} and observe the changes in \(\text{HRecall}@b@1000\) on industrial dataset, as shown in Fig. \ref{fig:scale}. Here we set \(K = K' = 1000\). The gray line represents the baseline performance under traditional beam search. Additionally, experiments with brute-force increasing the global beam size \(K\) under traditional beam search are included and marked with red triangles in the figure.

From Fig. \ref{fig:scale}, improvements in \(\text{HRecall}@b@1000\) are observed for \(b \in \{1, 2, 3\}\) as \(K^+\) increases. This indicates that the proposed path-level PRM can select an increasing number of optimal paths from the candidate set when \(K^+\) is enlarged. The gains at the second and third steps are particularly significant: for example, with \(K^+ = 6000\), \(\text{HRecall}@3@1000\) reaches 36.37\%, far exceeding the baseline value of 22.98\%. Importantly, the global beam size remains fixed at \(K = 1000\), meaning no additional computational resources for the decoder. The only increased computation lies in the lightweight path-level PRM. A detailed analysis of the computational cost of the path-level PRM is provided in Section \ref{sec:costs}.

In contrast, brute-force scaling of \(K\) yields only marginal metric improvements while significantly increasing the computational load of self-attention and cross-attention in the decoder. Therefore, the path-level PRM in \textsc{\modelname} successfully validates the \textbf{test-time scaling laws}.

\subsection{Analysis of Inference Efficiency (RQ6)} \label{sec:costs}
\vspace{-0.2cm}\begin{figure}[htbp]
    \centering
    \includegraphics[width=0.9\linewidth]{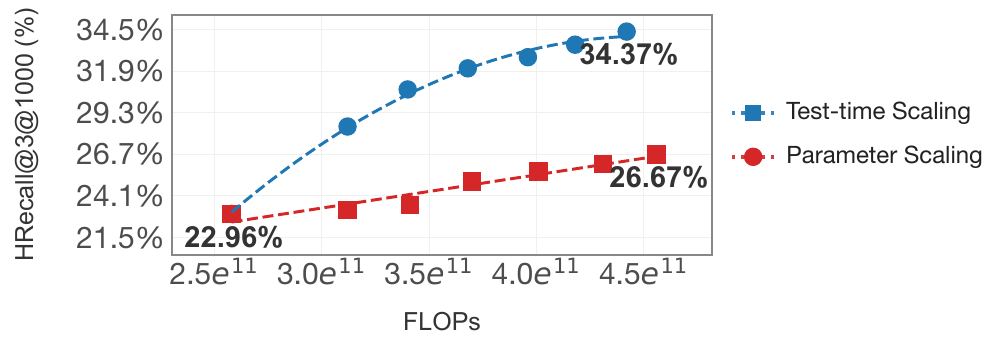}\vspace{-0.5cm}
    \caption{Comparison of test-time scaling with parameter scaling.}
    \label{fig:vs}\vspace{-0.3cm}
\end{figure}
\subsubsection{Test-time Scaling vs. Parameter Scaling.} Parameter scaling (increasing model size) is commonly used to scale recommendation models \cite{zhai2024actions,zhou2025onerec,zhou2025onerecv2}. We argue that test-time scaling offers greater inference efficiency. It can match the performance of parameter scaling with lower inference FLOPs.

Figure \ref{fig:vs} presents a comparison between test-time scaling and parameter scaling on our industrial dataset for generative recommenders. Both methods start from an identical base architecture with \(K=1000\). Test-time scaling involves keeping the model parameters fixed while introducing the PRM and gradually increasing \(K^+\) to improve performance. Parameter scaling fixes the beam size  but increases model size. The horizontal axis shows the inference FLOPs, while the vertical axis shows \(\text{HRecall}@3@1000\).

Our findings indicate that test-time scaling is more efficient. It achieves better performance at equal FLOPs, or reaches the same performance with fewer FLOPs. This demonstrates the superior efficiency of test-time scaling over enlarging model parameters.

\subsection{Hyper-Parameter Analysis (RQ7)}
\begin{figure}[htbp]
    \centering
    \includegraphics[width=\linewidth]{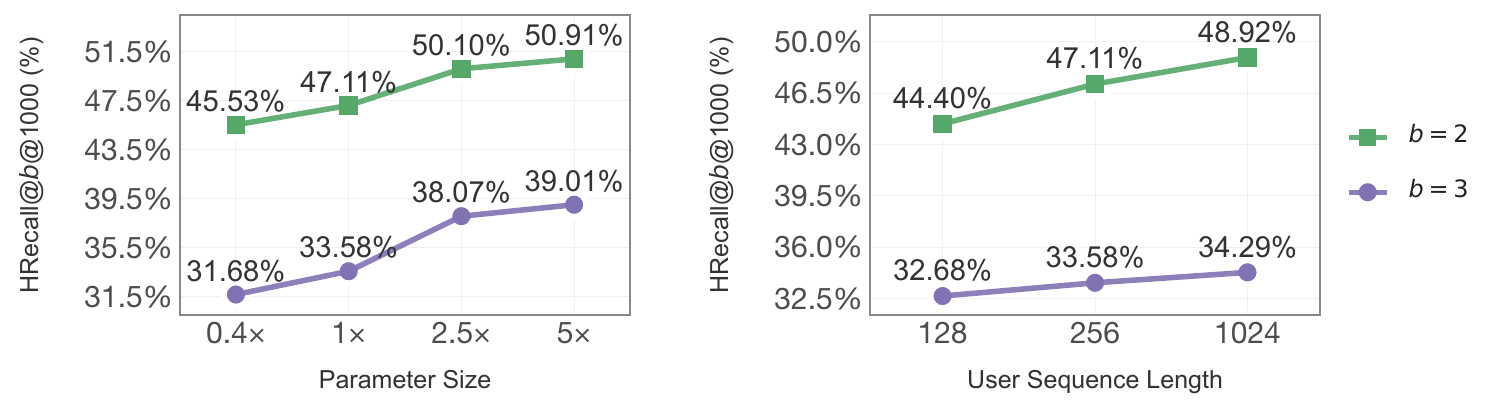}
    \caption{Impact of the model parameter size and the length of user sequence.}
    \label{fig:hyp}
\end{figure}
\subsubsection{Model Parameter Size.} We study the impact of model scale by analyzing the difference in \(\text{HRecall}@b@1000\) (\(b\in\{2,3\}\)) across various parameter sizes on the industrial dataset. As illustrated in Fig. \ref{fig:hyp}, model performance improves consistently with increased parameter size. This demonstrates that the proposed method scales effectively with model capacity, suggesting the advantage of our PRM mechanism in generative recommenders.

\subsubsection{User Sequence Length. } In the industrial dataset, we investigate the impact of the length of user sequences, shown in Fig. \ref{fig:hyp}. Model performance improves steadily as the sequence length increases. This can be attributed to the fact that the encoder representations are utilized by both the downstream generative task and the reward task. Richer information from the encoder effectively enhances recommendation performance.

\section{System Deployment}
We deployed the version with \(K^+=4000\) online. To keep it lightweight, we use only one PRM block. All queries are processed in parallel during cross-attention to maximize GPU utilization. We reduce the number of attention heads in the PRM to one-quarter of that in the main generation module. During inference, the decoder generates a larger candidate set (\(K^+=4000\)) compared to the original target size (\(K=1000\)), which increases the computational load for Top-K operations on the GPU. We mitigate this by introducing Radix Top-K optimization \cite{li2025radik}, which significantly speeds up top-k selection, resulting in lower latency for this step even with the expanded candidate set. In summary, with the aforementioned optimizations, the total parameter size increases by only 15\% compared to the version without PRM, while the inference latency increases by only 10\%.

\section{Related Work}
\subsection{Generative Recommendation}
Generative recommendation models perform autoregressive generation on user sequences to predict the next items a user will interact with. Research in this domain can be broadly categorized into three main areas: model architecture, item tokenization, and reward mechanism. 
Regarding model architecture, TIGER \cite{rajput2023recommender} innovatively adopts a T5-based encoder-decoder framework for sequential recommendation; HSTU \cite{zhai2024actions} further validates the scalability of decoder-only architectures; OneRec \cite{zhou2025onerec} proposes an end-to-end architecture to replace traditional cascaded recommenders; OneRec V2 \cite{zhou2025onerecv2}, introduces a lazy decoder structure to improve computational efficiency.
For item tokenization, common methods for generating Semantic IDs include RQ-VAE \cite{lee2022autoregressive}, Residual K-means \cite{luo2025qarm}, PQ \cite{jegou2010product}, and FSQ \cite{mentzer2023finite}.
Furthermore, several works introduce reward mechanisms to align with specific preferences: S-DPO \cite{chen2024softmax} enhances alignment with personalized ranking by improving the selection of preference data; OneLoc \cite{wei2025oneloc} enables the estimation of an item's potential GMV; Rec-R1 \cite{lin2025rec} directly optimizes the model using feedback from a fixed, black-box recommender system.

However, the exposure bias problem in NTP optimization is neglected in existing works. To address this, we are the first to propose a process reward mechanism for generative recommendation. It serves not only as a novel reward strategy but also as a new test-time scaling paradigm, yielding promising results.

\subsection{Process Reward Model}
In LLMs, test-time scaling based on process reward modeling has been shown to significantly enhance performance on mathematical reasoning or multi-step decision-making tasks. \citet{lightman2023let} successfully demonstrates that during inference, a per-step reward from a Process Reward Model (PRM) is more effective than the sparser reward from an Outcome Reward Model (ORM). Math-Shepherd \cite{wang2024math} significantly improves performance by re-ranking the solution steps generated by LLMs for mathematical problems. \citet{setlur2024rewardingprogressscalingautomated} propose Process Advantage Verifiers (PAVs) to quantify the quality of individual steps in multi-step reasoning. \citet{zhang2025entropyregularizedprocessrewardmodel} innovatively introduces an entropy-regularized process reward model. \citet{snell2024scalingllmtesttimecompute} confirms the feasibility of test-time scaling for LLMs, showing that a smaller language model, with additional inference-time computation, can surpass a model with 14 times more parameters on specific tasks. ThinkPRM \cite{khalifa2025processrewardmodelsthink} further optimize the PRM supervision paradigm. However, in the context of recommendation tasks, applying test-time scaling to improve the performance of generative recommendation has not yet been explored.

\section{Conclusion}
In this work, we addressed the critical challenge of Semantic Drift in generative recommendation by introducing \textsc{\modelname}, a framework powered by Process Reward Models. By providing dense, step-by-step feedback during the hierarchical generation of Semantic IDs, \textsc{\modelname} effectively mitigates error accumulation and exposure bias.  Our core innovation—the \textbf{Path-level PRM} coupled with \textbf{PRM-guided Beam Search}—not only enhances recommendation precision but also unlocks powerful \textbf{Test-Time Scaling Laws}. This allows for smaller models to outperform larger ones at a fraction of the serving cost.  

Experimental results across public benchmarks and real-world industrial environments confirm its effectiveness.
Online A/B test results prove their practical value in industrial scenarios.
We believe that the paradigm of test-time scaling through process supervision offers a promising and efficient direction for the next generation of industrial-scale recommender systems.

\bibliographystyle{ACM-Reference-Format}
\bibliography{sample-base}










\end{document}